\documentclass{article}  
\usepackage{breckenridge}
\frompage{000} \topage{000}                                              

\newcommand{\nc}{\newcommand}
\nc{\scm}{\sqrt{s_{\rm NN}}}
\nc{\equ}{{\rm equ}}
\nc{\bb}{{\rm {\bf b}}}
\nc{\bs}{{\rm {\bf s}}}
\nc{\pt}{p_{\rm T}}
\nc{\mt}{m_{\rm T}}
\nc{\pL}{p_{\rm L}}
\nc{\ET}{E_{\rm T}}
\nc{\Nch}{N_{\rm ch}}
\nc{\Nc}{N_{\rm coll}}
\nc{\Np}{N_{\rm part}}
\nc{\Atanh}{{\rm Atanh}}
\nc{\Asinh}{{\rm Asinh}}
\nc{\Acosh}{{\rm Acosh}}
\nc{\se}{\section}
\nc{\suse}{\subsection}
\nc{\beq}[1]{\begin{equation}\label{#1}}
\nc{\eeq}{\end{equation}}
\nc{\bea}[1]{\begin{eqnarray}\label{#1}}
\nc{\eea}{\end{eqnarray}}
\nc{\bce}{\begin{center}}
\nc{\ece}{\end{center}}
\nc{\bit}{\begin{itemize}}
\nc{\eit}{\end{itemize}}
\nc{\bmp}{\begin{minipage}}
\nc{\emp}{\end{minipage}}

\nc{\la}{\langle}       
\nc{\lla}{\left \langle}
\nc{\ra}{\rangle}       
\nc{\rra}{\right \rangle}

\def\rightmark{Expansion rates at RHIC}
\def\leftmark{Peter F. Kolb}

\title{Expansion rates at RHIC}

\authors{
{Peter F. Kolb}\\[2.812mm]
{\normalsize 
Department of Physics and Astronomy \\ 
SUNY Stony Brook\\
Stony Brook, NY, 11780
}}
\abstract{
A detailed description of the temporal evolution of 
the thermodynamic fields in heavy ion collisions is presented
within a hydrodynamic framework. 
Particular attention is devoted to the evolution of the collective flow
fields and  their space-time gradients.
}

\begin{document}
\maketitle
%
%
%
%
%
%
\section{Introduction and Motivation}
\label{intro}

Experimental data from RHIC \cite{QuarkMatters}, 
in particular the systematic analysis of an\-iso\-tro\-pies in the particle spectra, 
indicate that at high center of mass energies the collision region
rapidly thermalizes and for the major part of its evolution subsequently 
expands according to the laws of ideal hydrodynamics
(for a recent review that collects the arguments and
references which support this hydrodynamic scenario see~\cite{KH03}).
This is an important finding by itself as apparently the generated system 
spends some time at temperatures above the anticipated QCD transition
temperature and evolves according to the pressure, pressure-gradients
and thus equation of state of deconfined strongly interacting matter.
Furthermore  the hydrodynamic evolution can be used 
to shed light even onto some microscopic scattering phenomena,
such as signals from hard probes which propagate through the
hydrodynamically evolving 'background' fields and radiate energy, 
as well as the late freeze-out dynamics of the reaction.
Such probes are significantly dependent on the evolution of the 
macroscopic fields such as temperature and particle density, but also on the
collective flow velocities and their gradients. 
We will here give a detailed study of the evolution of these fields. First we 
investigate a simple analytically accessible parametrization of the flow field 
and its divergence. Then we will discuss the hydrodynamic evolution of
thermodynamic fields in a full hydrodynamic calculation to return later to 
the collective flow fields and compare to the initial analytic example.


\section{Expansion rate and dilution rate}
\label{sec:rates}  

Ideal hydrodynamics incorporates the continuity equations for conserved charges
$\partial _\mu n^\mu =0$ where $n^\mu= n\, u^\mu$ is the current associated
with the conserved charge whose density distribution is given by $n(x)$  
and $u^\mu(x)$ is the four-velocity of the collective flow field. 
In ideal hydrodynamics the entropy density is such a conserved quantity.
Slightly reshaping the continuity equation leads to the relation
\begin{equation}\label{equ:exprate}
\partial_\mu u^\mu = - \frac{1}{n} u^\mu \partial_\mu \, n \,.
\end{equation} 
The expression $\partial_\mu u^\mu$ is commonly referred to as {\em expansion rate}.
Considering the trivial condition of vanishing collective flow $u^\mu=(1,0,0,0)$ as is the 
case in the center of the system due to symmetry, the expansion rate is simply given
as $\partial_\mu u^\mu = - \dot{n} / {n}$, the {\em dilution rate}.
Assuming furthermore a power-law decay of the density $n \sim \tau^{-\alpha}$ as 
is the case for the three-dimensional Hubble expansion of the Universe ($\alpha =3$)
or the one-dimensional longitudinal Bjorken expansion ($\alpha = 1$)
the expansion rate is given by $\partial_\mu u^\mu = \alpha / \tau$
($\alpha$ is referred to as {\em expansion parameter}).
Let us now investigate a more general situation, namely a longitudinal Bjorken flow 
field $v_z= \tanh \eta = z/t $ 
with a transverse flow component given as 
$\gamma \,  v_T= \tanh (\xi r)$. 
We will later see that the hydrodynamic calculations in fact suggest such a flow profile
at freeze-out with $\xi \approx 0.07$~fm$^{-1}$.
It is quickly shown that this flow field leads to an expansion rate
\begin{equation}\label{equ:dmuumusource}
\partial_\mu u^\mu =
\tau^{-1}  \cosh (\xi r)
+
\xi \cosh (\xi r) + r^{-1} \sinh (\xi r)\,,
\end{equation}
which in the limit $(r, z) \rightarrow 0$ 
leads to 
$\partial_\mu u^\mu \rightarrow 1/\tau + 2 \xi$ as discussed in \cite{TW02}.
Keeping the radial dependence in the expression one finds that the expansion rate
increases with radial distance from 
0.207 fm$^{-1}$ at $r=0$ to $0.247$  fm$^{-1}$ at $r=10$~fm 
(assuming $\tau=15$~fm/$c$, a typical fireball lifetime). 
This will become important later in this study. 
In the case of such a  non-zero transverse flow profile the spatial part of 
the four product on the right hand side of Eq. (\ref{equ:exprate}) does not 
vanish and the expansion rate is no longer identically equal to the dilution rate. 
This becomes apparent by assuming a boost invariant distribution of the conserved charge 
(which means that the charge density only depends on the proper time $\tau=\sqrt{t^2-z^2}$ and 
not on the space-time rapidity $\eta$) 
and describing its dilution locally in terms of a power law such that 
$n(r,\tau) = n_0(r) (\tau_0/\tau)^\alpha$.
It then quickly follows from Eq. (\ref{equ:exprate}) that
\begin{equation}\label{equ:expvsdilut}
\tau \, \partial_\mu u^\mu = \gamma \left( \alpha - \tau \, v_r \frac{\partial_r n_0(r)}{n_0(r)} \right)\,.
\end{equation}
As the density is expected to drop with increasing radial distance the last term 
in the brackets is negative and gives a positive correction to $\alpha$
 as does the preceding $\gamma$ factor. 
It thus follows that in general  the expansion rate $\partial_\mu u^\mu$ 
is greater or equal to the local dilution rate $\alpha / \tau$.


\section{Time evolution of thermodynamic fields in central collisions}
\label{evolution}

We now investigate the evolution of the thermodynamic fields using a 
hydrodynamic model \cite{KSH00} which explicitly assumes boost-invariance
throughout the evolution.  
The initial field configuration in the transverse plane is determined by an 
optical Glauber-parametrization \cite{KHHET01} and the 
parameters of the calculation, 
namely the equilibration time $\tau_\equ$ and the constant $s_\equ$ which 
relates the geometrical density to the initial entropy density
are obtained by fitting to preliminary particle spectra from central 
Au+Au collisions  with $\scm = 200$~GeV.
Good agreement is achieved with 
$s_\equ=110$~fm$^{-3}$ at $\tau_\equ=0.6$~fm/$c$
\cite{KR03}, to be used in the following.

\begin{figure}[hbtp]
\vspace*{-.1cm}
\begin{center}
\begin{minipage}[t]{4.6cm}
    \epsfxsize 4.6cm \epsfbox{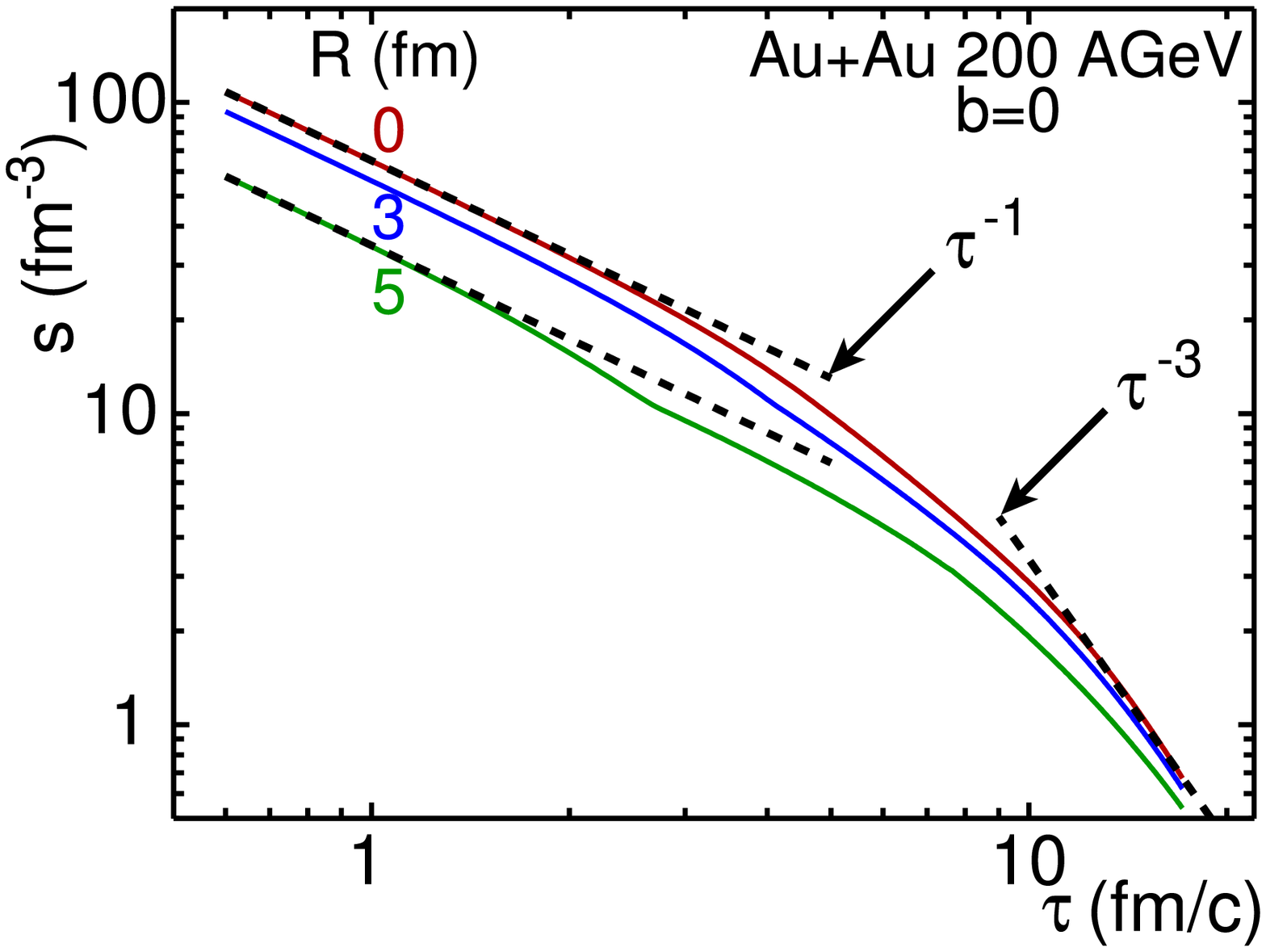}
\end{minipage}
    \hspace*{.3cm}
\begin{minipage}[t]{4.6cm}
   \epsfxsize 4.6cm \epsfbox{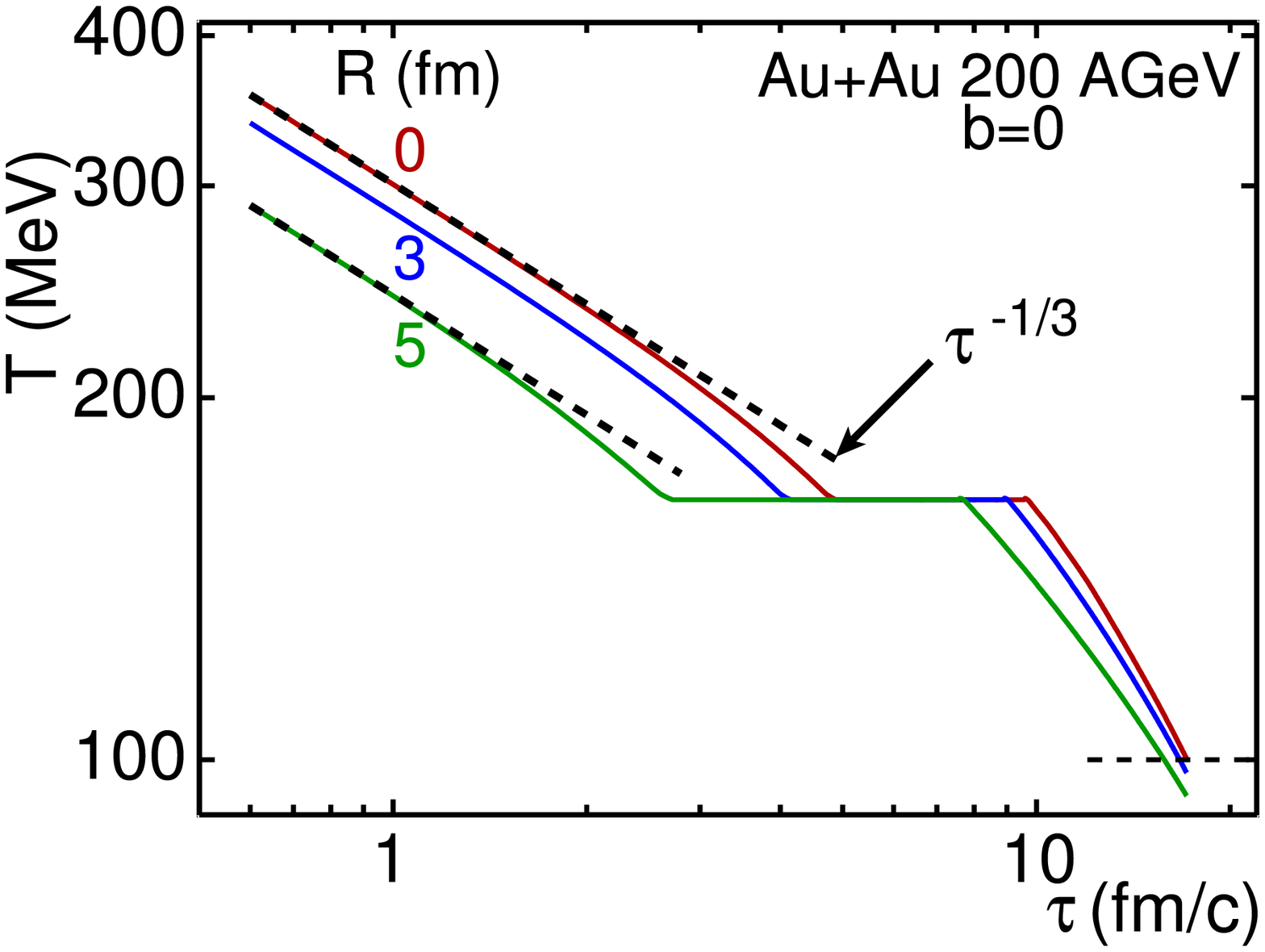}
\end{minipage}
  \end{center}
%
%
\vspace*{-.8cm}
\caption[]{
	Evolution of the entropy density (left) and the temperature (right)
	for central Au+Au collisions at $\scm=200$~GeV. Shown
              are the densities and temperatures at distances $R=0$, 3 and 5~fm
	from the center of the system.
}
%
%
\vspace*{-.2cm}
\label{fig:evosevoT}
\end{figure}

{\bf Evolution of entropy and temperature:}
We first study the time evolution of the entropy density 
and the temperature at the center of the system and
at 3 and 5 fm radial distance (Fig. \ref{fig:evosevoT}).
Initially the entropy decays with a $\tau^{-1}$ power law due
to the initial longitudinal Bjorken expansion 
(see the dashed lines). 
As time evolves, the transverse pressure gradients lead to a collective
radial expansion and thus a transition from the initial one-dimensional
to a full three dimensional expansion in the late stage. 
At late times the dashed line indicates a decay $\sim \tau^{-3}$. 

The equation of state supplemented to the hydrodynamic equations 
of conservation of energy and momentum features a strong first order phase
transition at a critical temperature of $T_c=164$~MeV. 
Above that temperature the system acts like an ideal gas with 
$c_s^2 = \partial p/\partial e = 1/3$ and $e \sim T^4$. 
For a one-dimensional Bjorken expansion the energy density decays 
as $e \sim \tau^{-(1+c_s^2)}$ and thus $T \sim \tau^{-1/3}$.
The figure shows that this behavior is neatly followed until just before the system 
hits the mixed phase where the temperature remains unchanged while the
system boils off its latent heat. 
Further cooling appears then in the resonance gas phase below~$T_c$.

\begin{figure}[hbtp]
\vspace*{-1mm}
\begin{center}
\begin{minipage}[t]{4.6cm}
    \epsfxsize 4.6cm \epsfbox{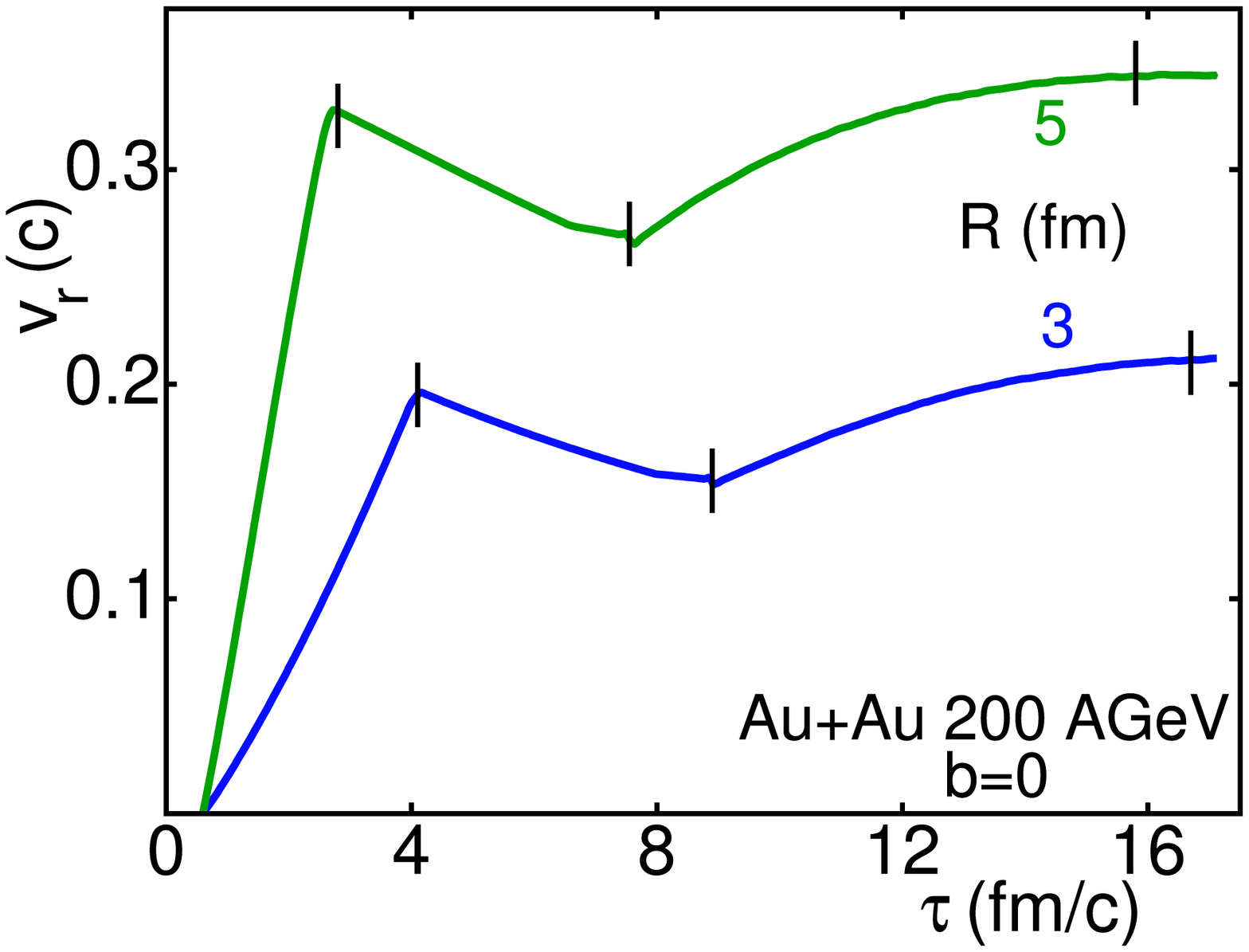}
\end{minipage}
    \hspace*{.3cm}
\begin{minipage}[t]{4.6cm}
   \epsfxsize 4.6cm \epsfbox{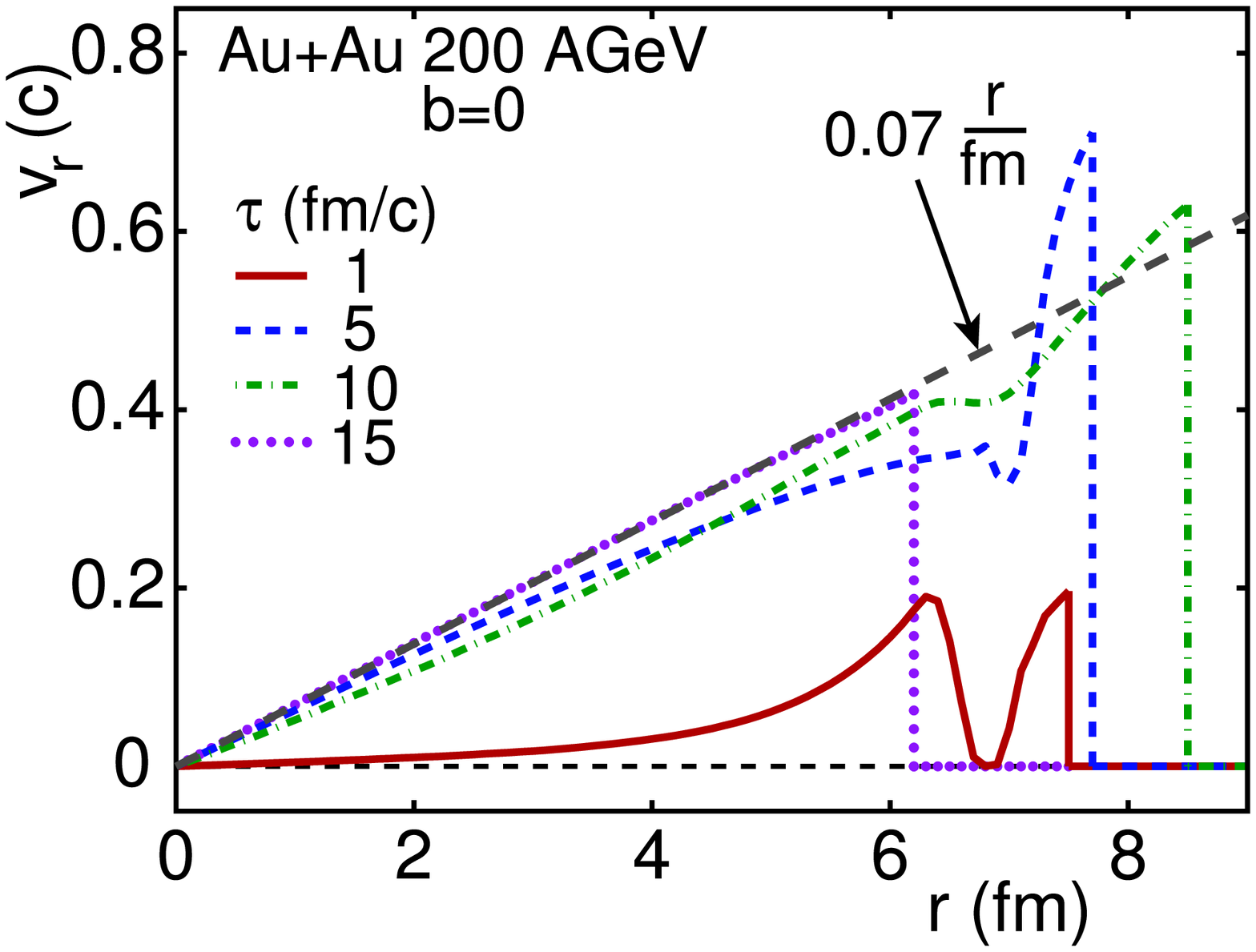}
\end{minipage}
\end{center}
%
%
\vspace*{-.8cm}
\caption[]{
	Evolution of the radial velocity field. The left plot
	shows the velocity as a function of time at positions 3 and 5 fm 
   	away from the center of the system. 
              The right plot shows
	the radial flow profile as a function of the radial distance
	at 4 different times together with a linear relation 
              $v_r = \xi r$ with $\xi = 0.07$~fm$^{-1}$.
}
%
%
\vspace*{-.2cm}
\label{fig:evovrad}
\end{figure}

{\bf Evolution of radial velocity:} 
Now we consider the collective radial velocity which is generated through the
act of the pressure gradients in the system. 
The left plot of Fig. \ref{fig:evovrad} shows the radial flow velocity at distance 3 and
5 fm from the fireball center. 
As pressure gradients are larger at 5 fm than at 3 fm, the acceleration is stronger and
thus the rise of $v_r(\tau)$. 
When the system reaches the mixed phase,  
the acceleration ceases and the velocity locally drops 
as matter continues to move radially outward, thus transporting the slower
fluid cells from the interior to larger radial distances.
Only as the matter cools into the hadronic phase,
pressure gradients pick up again  and more radial flow is generated.

The right hand plot of Fig. \ref{fig:evovrad} shows the radial flow field as a function
of radial distance at four different times. 
The generated flow field  1~fm/$c$ after impact has an interesting structure,
showing a non-monotonous behavior due to the mixed phase
where no collective flow develops. 
As time evolves the field rapidly adopts a linear behavior which is largely
preserved for the rest of the evolution, and can very well be described by
$v_r(r)= \xi\,r$ ($\approx \tanh (\xi \, r)$) with $\xi = 0.07$~fm$^{-1}$.
This 'static' flow field is at the heart of the successful description 
of a large number of observables in terms of 
the popular  'blast wave' parametrization \cite{SSH93}.

{\bf Evolution of dilution- and expansion rates:} 
We now return to the dilution- and expansion rates in the full hydrodynamic calculation. 
\begin{figure}[hbtp]
\vspace*{-1mm}
\begin{center}
\begin{minipage}[t]{4.8cm}
    \epsfxsize 4.8cm \epsfbox{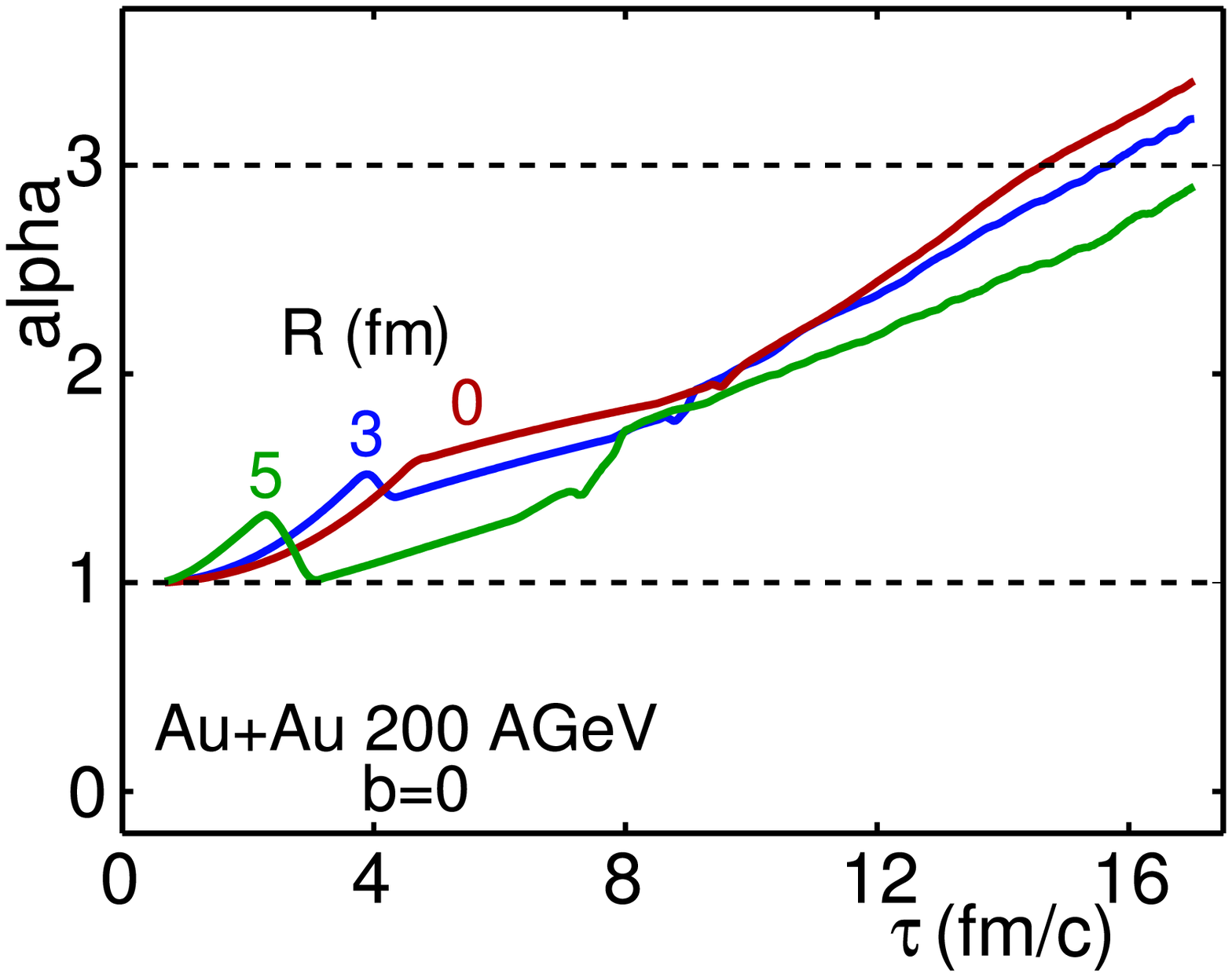}
\end{minipage}
    \hspace*{.1cm}
\begin{minipage}[t]{4.9cm}
   \epsfxsize 4.9cm \epsfbox{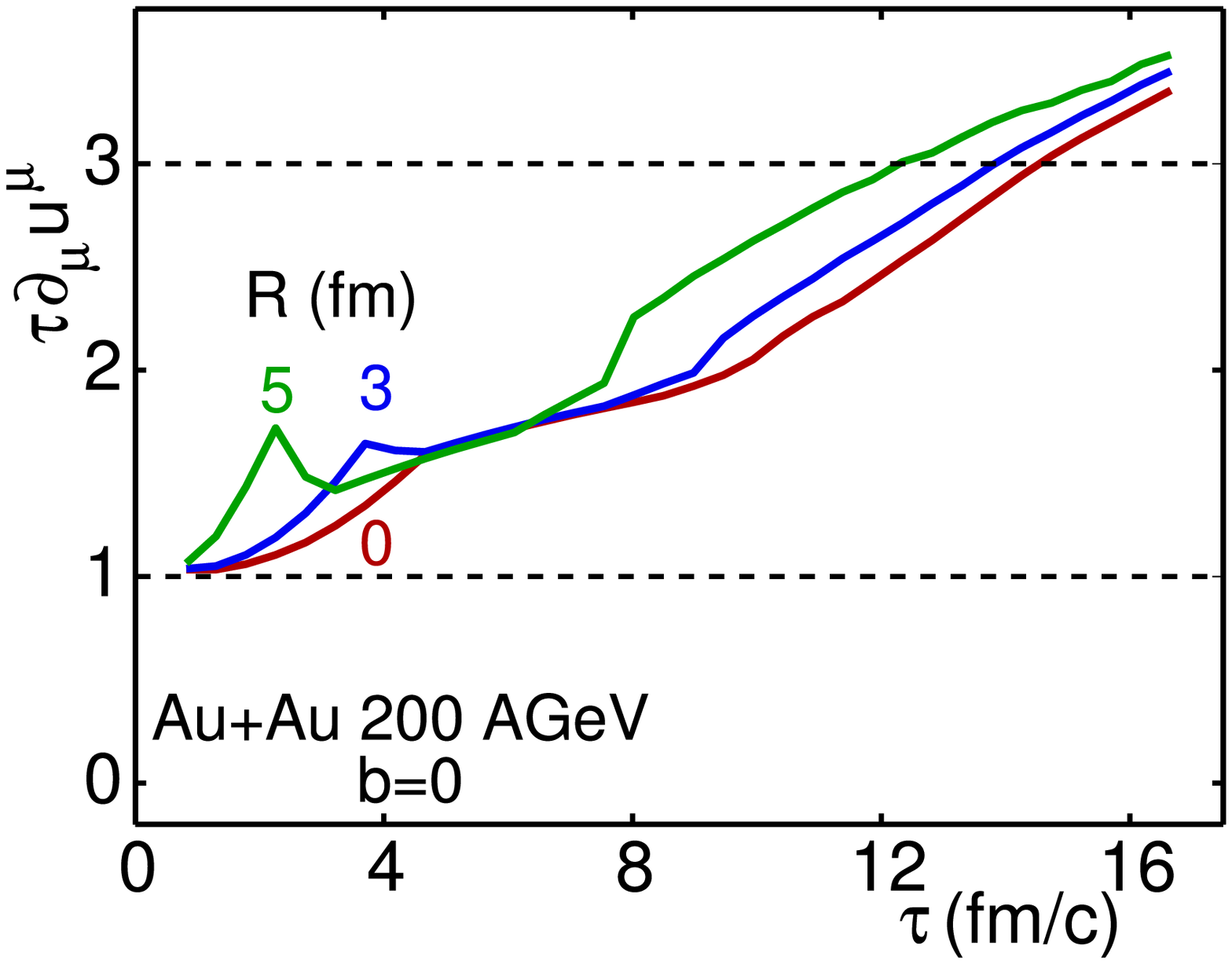}
\end{minipage}
\end{center}
%
%
\vspace*{-.8cm}
\caption[]{
	Evolution of the local expansion parameter $\alpha$
              and the product of expansion rate and time, $\tau \, \partial_\mu u^\mu$ 
              at distances $R=0$, 3 and 5~fm from the center.
}
%
%
\vspace*{-.3cm}
\label{fig:evoexpansion}
\end{figure}
Fig. \ref{fig:evoexpansion} shows the evolution of the local expansion parameter
$\alpha = -\frac{\tau}{s} \frac{\partial s}{\partial \tau} 
            = -\frac{\partial \log s}{\partial \log \tau}
$
in the left panel.  
It is clearly seen how the initial one-dimensional motion with
$\alpha=1$ becomes more three-dimensional and just reaches $\alpha =3$ at 
the time of decoupling (assuming freeze-out at $T_d=100$~MeV).
Note the influence of the radial position during late stages of the expansion. 
At larger radial distances the dilution rate is smaller,  as supplies from the 
interior are constantly streaming  outward.

The right panel of the figure shows the evolution of the product of expansion rate 
and time, $\tau \, \partial_\mu u^\mu$. 
Note that  with increasing the radial distance we get an opposite behavior as 
exhibited by the dilution rate. 
As demonstrated by the analytical calculation before, the expansion rate 
increases with radial distance.
The sign of the difference is understood in terms of Eq. (\ref{equ:expvsdilut}).


\section{Time evolution of anisotropies in non-central collisions}
\label{anisoevolution}

\begin{figure}[hbtp]
\vspace*{-1mm}
\begin{minipage}[t]{4.8cm}
    \epsfxsize 4.8cm \epsfbox{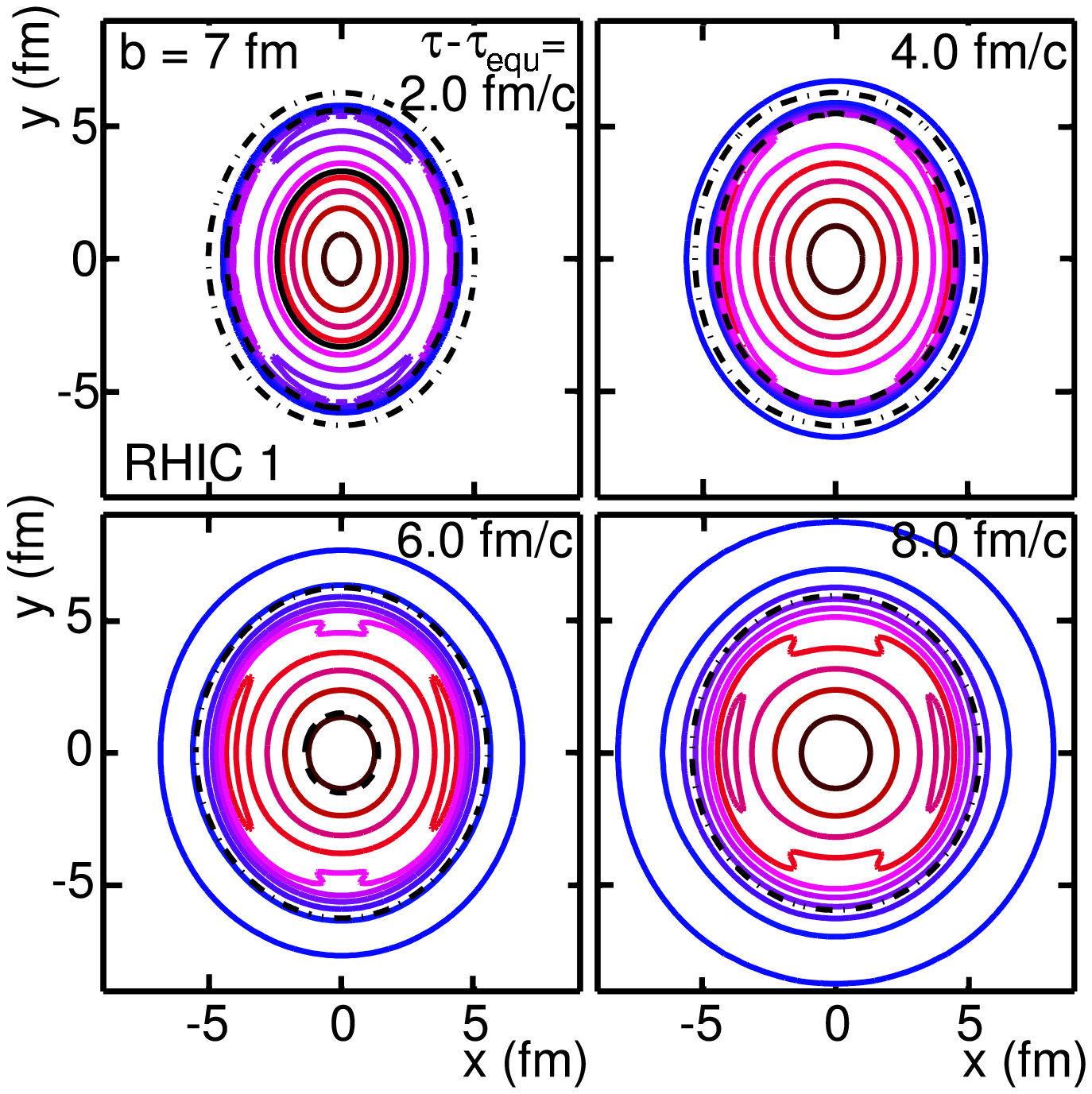}
\end{minipage}
    \hfill
\begin{minipage}[t]{6.5cm}
   \epsfxsize 6.5cm \epsfbox{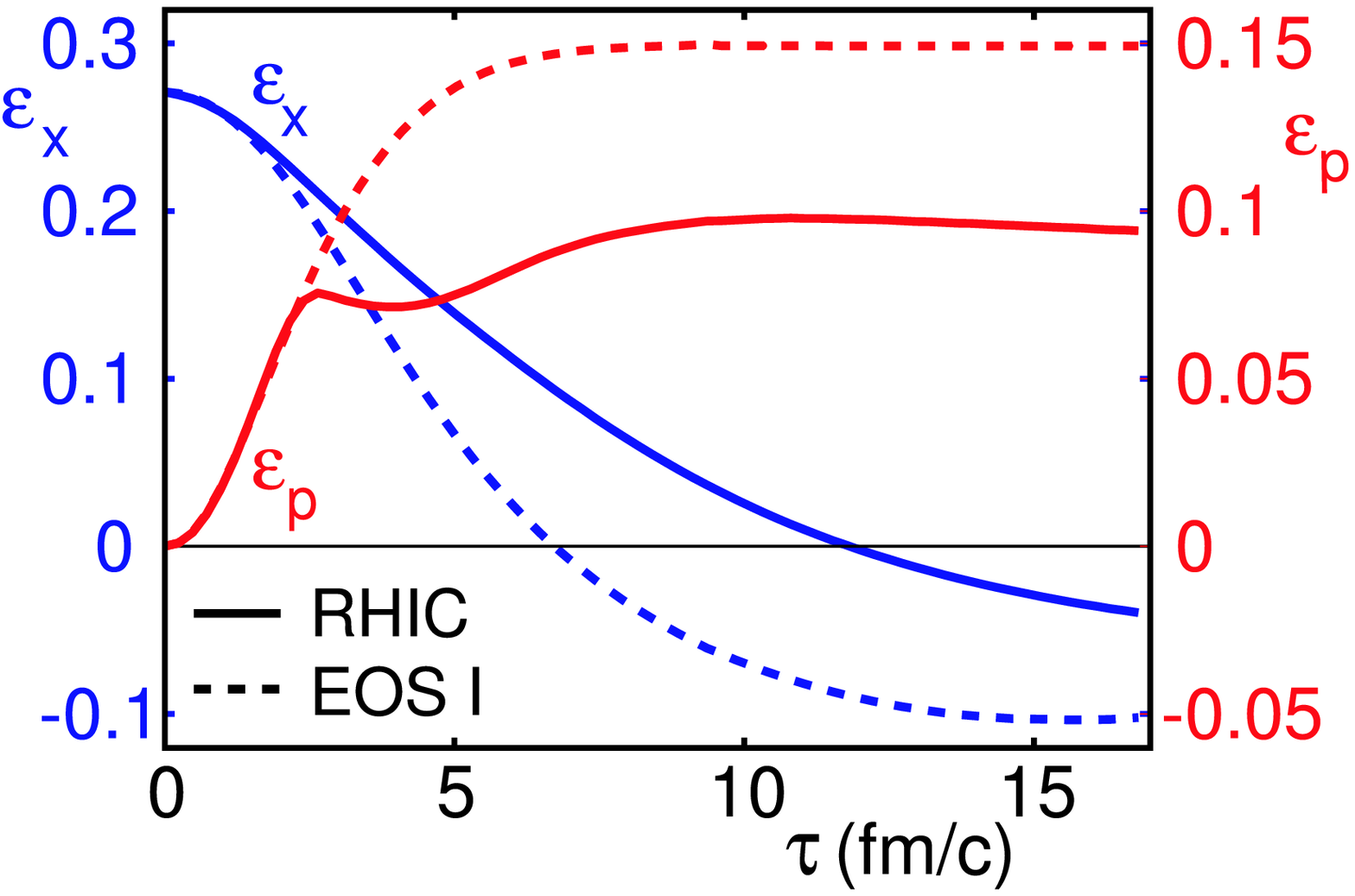}
\end{minipage}
%
%
\\[-.8cm]
\caption[]{Contours of the energy density in the transverse plane at four
 	different times during the systems' evolution. The figure
	to the right shows the time evolution of the spatial eccentricity
	$\epsilon_x$ and the momentum anisotropy $\epsilon_p$, for a system
	with (RHIC 1) and without (EOS I) a phase transition.
}
%
%
\vspace*{-.2cm}
\label{fig:evoanisotropies}
\end{figure}

Observables from non-central collisions deliver the strongest arguments
for rapid thermalization and a subsequent hydrodynamic expansion of the
reaction region \cite{hydrov2}. 
The geometric anisotropy in non-central collision $\epsilon_x$ results in 
anisotropic pressure gradients which in turn lead to anisotropies in the
collective flow field. Those can be characterized in terms of $\epsilon_p$, 
the asymmetry of the transverse part of the energy momentum tensor \cite{KSH00}.
This flow anisotropy will finally be observable as an azimuthal dependence
of particle emission (i.e. elliptic flow) \cite{Ollitrault92}.
From the left part of Fig. \ref{fig:evoanisotropies}, which shows
contours of constant energy density in the transverse plane at different times, 
it can be seen how the initial spatial eccentricity quickly diminishes.
This becomes more quantitative in the right part of the figure where 
the evolution of $\epsilon_x$ and $\epsilon_p$ for
a system with (RHIC 1) and without (EOS I) a first order phase transition 
is shown.
The large initial geometric eccentricity leads to a rapid generation 
of momentum anisotropy, which in turn reduces the spatial eccentricity, 
and thus cuts off the  further generation of momentum anisotropy.
During the soft phase of the transition pressure gradients vanish (and thus also 
their anisotropies), which leads to an even earlier saturation of the momentum 
anisotropy. 
For systems at RHIC most of the anisotropy is thus generated in the plasma stage
of the fireball.


\section{Summary and Relevance}
\label{summary}

We have given a detailed account on the evolution of the thermodynamic fields 
in systems created at RHIC within a hydrodynamic framework. 
This information is relevant for the microscopic dynamics at different levels. 
At early stages, hard probes are to be transported over the 
varying thermodynamic background fields to calculate their energy loss. 
At late stages, the scattering rates in the resonance gas needs to be compared
to the macroscopic expansion and dilution rates to investigate the overall 
reliability of the idealized macroscopic approach which requires a 
'slow' (adiabatic) evolution and strong rescattering.
Furthermore there are sensitive probes of the density around decoupling
such as light nuclei, whose formation probability favors a large
density, but their survival rate increases with decreasing density. 
Resonances reconstructed from a hadronic decay 
channel have similar information content \cite{resonances}.
In summary,  essentially all experimental observables are influenced by the temporal 
evolution of the thermodynamic fields presented here either as the signal
is sensitive to all stages of the evolution and thus integrates the temporal history
of the system 
or as it results from a particular stage of the collision with densities and 
their gradients giving the relevant physical background.
\\[-2mm]
 
{\bf Acknowledgments:}
This research was supported in part by the U.S. Department of Energy
under Grant No. DE-FG02-88ER40388. Support from the 
Alexander von Humboldt Foundation through a Feodor Lynen 
Fellowship is gratefully acknowledged.


\vfill\eject
\end{document}